\font\titlefont=cmbx10 scaled\magstep1
\magnification=\magstep1
\null
\rightline{SISSA xx/96/EP}
\vskip 1.5cm
\centerline{\titlefont Yang--Mills vacuum structure}
\centerline{\titlefont  and quantum gravity}
\smallskip
\vskip 1.5cm
\centerline{\bf S. Odintsov \footnote{$^*$}{\tt 
sergei@ecm.ub.es}}             
\smallskip
\centerline{Inst.of Theoret.Physics and Center for Theoret.Sciences}
\centerline{Leipzig University, Augustusplatz 10,D-04109 Leipzig, Germany}
\centerline{ and Tomsk Pedagogical University, 634041 Tomsk,Russia}
\bigskip\smallskip
\centerline{\bf R. Percacci \footnote{$^{**}$}{\tt 
percacci@neumann.sissa.it}}
\smallskip
\centerline{International School for Advanced Studies, Trieste, 
Italy}
\centerline{via Beirut 4, 34014 Trieste, Italy}
\centerline{and}
\centerline{Istituto Nazionale di Fisica Nucleare, 
Sezione di Trieste}
\vskip 1.8cm
\centerline{\bf Abstract}
\smallskip
\midinsert
\narrower\narrower\noindent
Using renormalization group methods, we calculate the
derivative expansion of the effective Lagrangian for a
covariantly constant gauge field in curved spacetime.
Curvature affects the vacuum; in particular
it could induce phase transitions between different vacua.
We also consider the effect of quantum fluctuations
of the metric, in the context of a renormalizable $R^2$
theory. In this case the critical curvature depends on
the gravitational coupling constants.
\endinsert
\vskip 1cm
\vfil\eject
\leftline{\bf 1. Introduction}
\smallskip
\noindent
The one-loop effective Lagrangian for a gauge theory
contains a term, called the chromomagnetic potential,
describing the response of the system to a covariantly constant
chromomagnetic field (by this we mean the space component of
the field strength; the gauge field need not necessarily
be the gluon. In fact this kind of analysis has been applied to
electroweak theory.)
It has been shown that the chromomagnetic potential
can have a non zero minimum; the corresponding state is then
called a chromomagnetic vacuum.
It was also realized that such a state could not be the true
vacuum of the theory: the effective potential has an imaginary part, due
to the presence of an unstable field mode.
There have been various proposals on how this state may be stabilized.

One may expect the presence of large chromomagnetic fields in the
early universe, when spacetime curvature was not zero. Hence it
seems natural to investigate the influence of (background)
curvature effects, or even quantum gravity effects, on the chromomagnetic
vacuum. In this paper we address this problem by calculating
the chromomagnetic potential for a covariantly constant gauge
field in curved spacetime, using renormalization group methods and
the derivative expansion. In the case of an SU(2) gauge field, we
show that curvature influences the position of the minimum of
the chromomagnetic potential. There is the possibility of
curvature--induced phase transitions from the nontrivial
phase (for negative or positive but small curvature)
to a phase with zero chromomagnetic field (for curvature above
a certain positive critical value). The v.e.v. of the chromomagnetic
field grows as curvature becomes more negative.

We then consider the contribution of a renormalizable $R^2$--type
quantum gravity to the chromomagnetic potential.
Again we will see that there is the possibility of having 
phase transitions, depending now on the background curvature
as well as new parameters appearing in the effective action.

\leftline{\bf 2. Renormalization group improved effective Lagrangian}
\smallskip
\noindent
It is well-known (see for example [1] or, for a general review, [2])
that the renormalizable Lagrangian for a nonabelian gauge theory in 
curved spacetime is given by
$$
L=-{1\over4}G_{\mu\nu}^aG^{a\mu\nu}+a_1 R^2
+a_2 C_{\mu\nu\rho\sigma}C^{\mu\nu\rho\sigma}
+a_3 G\ ,\eqno(1)
$$
where $G_{\mu\nu}^a=\nabla_\mu A_\nu^a-\nabla_\nu A_\mu^a
+g f^a{}_{bc} A_\mu^b A_\nu^c$, $\nabla$ is the spacetime
covariant derivative, $g$ is the gauge coupling constant,
$a_i$ are gravitational coupling constants, $C$ is the Weyl
tensor and $G$ is the Gauss--Bonnet term.

This theory is known to be asymptotically free in flat spacetime [3]
(for a discussion using the background field method see [4]).
In curved spacetime the renormalization of $g$ is the same as in flat space
[1] and hence the theory is asymptotically free in the gauge coupling
also in curved spacetime [1,2].
The last three terms represent the Lagrangian of the external 
gravitational field which should be added in order to have multiplicative
renormalizability  in curved spacetime [2].

Due to this multiplicative renormalizability we can
easily derive the following renormalization group (RG)
equation [2]:
$$
\left(\mu{\partial\over\partial\mu}
+\beta_i{\partial\over\partial\lambda_i}
-\gamma_A A_\mu^a{\delta\over\delta A_\mu^a}
\right)L_{\rm eff}(\mu,\lambda_i,A_\mu^a,g_{\mu\nu})=\,0
\ ,\eqno(2)
$$
where $\lambda_i=(g,a_1,a_2,a_3)$, $\gamma_A$ is the standard RG
gamma function for  the (background) field $A_\mu^a$.

We will discuss the RG improved effective Lagrangian obtained as the 
direct solution of the RG equation.
This technique is quite well-known from applications to the 
Coleman--Weinberg effective potential in flat [6] and curved [7] space, 
as well as the effective Lagrangian for a covariantly constant gauge field
[5,8,10]. Note however that the usual electroweak vacuum becomes unstable 
for large magnetic field [8] (for a discussion of how stability may be 
restored in the electroweak theory see [9]).

Working in the background field method we will use the standard relation
$$
\beta_g=g\gamma \eqno(3)
$$
and the one-loop beta function $\beta_g=-{b g^3\over(4\pi)^2}$,
$b>0$, where the value of the coefficient $b$ depends on the gauge group
under consideration.

In the solution of the RG equation (2) it is convenient to represent
the effective Lagrangian as follows:
$$
L_{\rm eff}=-{1\over4}G_{\mu\nu}^aG^{a\mu\nu} 
f_1(\mu,\lambda_i,A_\mu^a,g_{\mu\nu})+ 
L_{\rm ext}(\mu,\lambda_i,A_\mu^a,g_{\mu\nu})\ ,\eqno(4)
$$
where $f_1$ is a dimensionless function, $L_{\rm ext}$ is the vacuum
polarization (or quantum effective Lagrangian of the external fields)
and the background gauge field in $f_1$ and $L_{\rm eff}$ always appears
in some dimensionless combination.

Applying the RG equation to the effective action in the form (4)
one can solve it as follows:
$$
f_1(\mu,\lambda_i,A_\mu^a,g_{\mu\nu})=
\exp\left(-2\int_0^t\bar\gamma(g(t'))dt'\right)
f_1(\mu e^t,\bar\lambda_i(t),A_\mu^a,g_{\mu\nu})\ ,\eqno(5)
$$
where
${d\bar\lambda_i(t)\over dt}=
{\bar\beta_i(\bar\lambda_i(t))\over1+\bar\gamma(\bar g(t))}$,
$\bar\gamma={\gamma\over 1+\gamma}$ and the condition (3)
has been used for the effective coupling.
Below we will consider only the leading order approximation, where
$$
\bar\beta_i={\beta_i\over 1+\gamma}\approx \beta_i\ \ ;
\qquad\qquad
\bar\gamma={\gamma\over 1+\gamma}\approx \gamma \eqno(6)
$$
and the one loop beta and gamma functions should be used in (6).
For $L_{\rm ext}$ we get
$$
L_{\rm ext}(\mu,\lambda_i,A_\mu^a,g_{\mu\nu})=
L_{\rm ext}(\mu e^t,\lambda_i(t),A_\mu^a,g_{\mu\nu})\ .\eqno(7)
$$
where
${d\lambda_i\over dt}=\beta_i(\lambda(t))$.

Collecting (5) and (7) in (4) we get the RG improved effective Lagrangian
for YM theory in curved spacetime.

Working in leading-log approximation, one should use some boundary
condition to define the effective action at $t=0$.
As a boundary condition we will use the tree-level Lagrangian (1).
As a result the RG improved effective action has the following form:
$$
L=-{1\over4}{g^2(0)\over g^2(t)}G_{\mu\nu}^aG^{a\mu\nu}
+a_1(t) R^2
+a_2(t) C_{\mu\nu\rho\sigma}C^{\mu\nu\rho\sigma}
+a_3(t) G\ ,\eqno(8)
$$
where eqs. (3) and (6) have been used to rewrite $f_1$.
The effective gravitational coupling constants are known
for any renormalizable matter theory [2].

The next issue is the meaning of the RG parameter $t$.
Usually $t$ is taken as the logarithm of the
effective mass appearing in the one-loop effective action.
Another way (giving an equivalent result)
is to use the one loop effective action as a boundary condition
after solving the RG equation. Then $t$ is defined by the condition of
vanishing of the logarithmic term in the effective action.

The one-loop effective action in Yang--Mills theory
in curved spacetime may be easily found in the background field method
(in the gauge $\alpha=1$):
$$
\Gamma^{(1)}={1\over2}\ln\det{\cal D}_{\mu\nu}^{-1}
-\ln\det(-\hat\triangle)\ ,\eqno(9)
$$
where $A_\mu=g A_\mu^a T_a$, $G_{\mu\nu}=G_{\mu\nu}^a T_a$,
$\hat\nabla$ is the gauge and spacetime covariant derivative,
$\hat\triangle=\hat\nabla_\mu\hat\nabla^\mu$
and
$$
{\cal D}_{\mu\nu}=-g_{\mu\nu}\hat\triangle
+R_{\mu\nu}-2 G_{\mu\nu}\ .\eqno(10)
$$
Let us now turn to the discussion of the specific situation
where the general formalism will be used.
\smallskip
\leftline{\bf 3. The vacuum of SU(2) gauge theory in curved spacetime}
\smallskip
\noindent
We will now consider an SU(2) YM theory in curved spacetime.
The determination of the covariantly constant field will be
a straightforward generalization of the flat space condition:
$$
\hat\nabla^{\mu ab}G_{\mu\nu}^b=\,0\ .\eqno(11)
$$
The solution of this equation may be found in a normal coordinate expansion
where the first term will give the flat space solution and the
remaining ones give curvature corrections.
For the group SU(2) there are two invariant combinations of the curvatures:
$$
F={1\over 4}G_{\mu\nu}^aG^{a\mu\nu}\ \ ;
\qquad\qquad
Y={1\over 4}G_{\mu\nu}^aG^{a\ast \mu\nu}\ ,\eqno(12)
$$
so for the covariantly constant magnetic gauge field we get
$$
F={1\over2}H^2+O(R)\ .\eqno(13)
$$
Let us choose a background satisfying the condition
$R_{\mu\nu}={1\over4}R g_{\mu\nu}$. From the results of refs.[5,8]
we know that in flat space the effective mass is given by $gH$,
hence in flat space $t={1\over2}\ln{gH\over\mu^2}$ [5,8]
(this can be seen by diagonalizing the mass matrix or by direct proper-time
calculation). 

At the same time we know that in curved spacetime with vanishing
gauge field the effective mass is given by the curvature.
Hence a possible choice of the RG parameter $t$ is
$$
t={1\over2}\ln{{R\over4}+gH\over\mu^2}\ .\eqno(14)
$$
The RG improved effective Lagrangian for a magnetic field in
curved spacetime is then
$$
L_{\rm eff}=-{1\over2}{g^2(0)\over g^2(t)}H^2
+a_1(t) R^2
+a_2(t) C_{\mu\nu\rho\sigma}C^{\mu\nu\rho\sigma}
+a_3(t) G\ ,\eqno(15)
$$
where
$$
g^2(t)={g^2(0)\over 1+{11g^2(0) t\over 12\pi^2}}\ ;
\qquad a_1(t)=a_1(0)\ ;
\qquad a_3(t)=a_3(0)-{62 t\over120(4\pi)^2}\eqno(16)
$$
and $G=R^2/6$, $C_{\mu\nu\rho\sigma}=0$ (due to the choice of background).
Then
$$
L_{\rm eff}=-{1\over2}{g^2(0)\over g^2(t)}H^2
+\left(a_1+{1\over 6}a_3-{62 t\over 720(4\pi)^2}\right)R^2\ .\eqno(17)
$$

For $g$ small, if next-to-leading terms of order $g$ can be neglected, the
minimum of the effective potential (which is minus of effective
lagrangian) occurs for
$$
gH_{\rm min}=\mu^2\exp\left(-{24\pi^2\over 11g^2}\right)\ -R/4.
\eqno(18)
$$
This solution generalizes the (unstable) chromomagnetic vacuum
of ref.[5] due to the influence of the gravitational background.
Note that unlike to above case the symmetry breaking for scalar
potential in curved spacetime may occur more frequently (actually
already on tree level) [11].

Furthermore we see that there is a curvature-induced phase transition
with critical curvature
$$
R_c=4\mu^2\exp\left(-{24\pi^2\over 11g^2}\right)\ .
\eqno(19)
$$
For positive curvature above $R_c$ the ground state of the
theory is $H_{\rm min}=0$. For curvatures below $R_c$ the ground state
has a novanishing background magnetic field.
Negative curvatures increases the value of this background.

This picture is confirmed by numerical calculations for $g$
not very small. For example choosing $g=0.5$ and
using the dimensionless variables
${gH\over\mu^2}=\exp\left(-{24\pi^2\over 11g^2}\right)x$ and
${R\over\mu^2}=\exp\left(-{24\pi^2\over 11g^2}\right)y$,
for $y$=2.2, 1, 0, -0.5,
the minimum occurs for $x$=0.21, 0.46, 0.6, 0.66
respectively.
The critical curvature occurs for $y\approx 2.3$.

Note that we cannot study the region of large negative curvature due to the
appearance of an imaginary part in the potential.
It is possible that for some value of the curvature this imaginary
part is cancelled by curvature effects. However to check this conjecture
one would have to treat the external gravitational field exactly.
\medskip

\leftline{\bf 4. Vacuum of SU(2) gauge theory coupled to higher 
derivative gravity} 
\smallskip
\noindent
Finally, we discuss the modifications which will appear in the
presence of quantized gravity. For this we need a workable model
of quantum gravity. We will work within the context of higher
derivative quantum gravity (for an introduction see [2]);
it can be considered here as an effective theory of quantum gravity.

We will start from the Lagrangian written in the following form:
$$
L=-{1\over4}{g^2(0)\over g^2(t)}G_{\mu\nu}^aG^{a\mu\nu}
+{1\over2\lambda} C_{\mu\nu\rho\sigma}C^{\mu\nu\rho\sigma}
-{\omega\over3\lambda}R^2\ ,\eqno(20)
$$
where $\lambda$ and $\omega$ are gravitational coupling constants.
The Gauss--Bonnet term can be neglected in this context.

Proceeding as in the previous section and using the fact that the
renormalization of the gauge coupling is not affected by the
new terms, we obtain the RG improved effective Lagrangian as
$$
L=-{1\over2}{g^2(0)\over g^2(t)}H^2
-{\omega(t)\over3\lambda(t)}R^2\ ,\eqno(21)
$$
where $g^2(t)$ is given again by (16) and
$$
\lambda(t)={\lambda(0)\over 1+{139\lambda(0)t\over10(4\pi)^2}} ,\eqno(22)
$$
$$
(4\pi)^2{d\omega(t)\over dt}=
-{139\over10}\lambda(t)\omega(t)
-\lambda(t)
\left[{10\over3}\omega^2(t)+5\omega(t)+{5\over12}\right]\ .
$$
The gravitational effective couplings may be found in [2].
As mentioned above, there is no choice of $t$ which eliminates
the logarithms to all orders in a theory with many mass scales.
Moreover, it is not realistically feasible to diagonalize the mass
matrix in the theory with Lagrangian (1).
Assuming that the magnetic field is of the same order of
(or larger than) the curvature, in leading log approximation 
a reasonable choice for $t$ will be
$$
t={1\over2}\ln{gH+cR\over\mu^2}\eqno(23)
$$
with $c$ a constant of order one.
We set $c=1/4$ as in the preceding section.

The approximate effective Lagrangian may now be written as follows:
$$
L_{\rm eff}=-{1\over2}\left(1+{11g^2 t\over12\pi^2}\right)H^2
+{1\over3(4\pi)^2}\left[{10\over3}\omega^2+5\omega+{5\over12}\right]t R^2\ .
\eqno(24)
$$

Numerical analyses show that the results are qualitatively the same as
in the previous section. However, the value of the critical curvature
depends on the coupling constants. For example, using again $g=0.5$
for the values
$\omega=-0.02$ and $\omega=-6$ (which are the approximate critical
points of RG
equation for $\omega$ )
one finds critical curvature at $y=2.1$ and $y=0.2$ respectively.
Hence, the phase diagram of the magnetic vacuum is affected by the
initial values of the gravitational couplings.
\medskip
\leftline{\bf 5. Conclusions}
\smallskip
\noindent
In summary, we have investigated the effective Lagrangian for
a covariantly constant nonabelian gauge field in the presence
of background and quantized gravitational field, using the RG equation
and the adiabatic expansion. We have found that there are curvature-induced
phase transitions between different vacua.

There are some obviuos extensions of the present work.
First, it would be interesting to discuss these phenomena
treating the gravitational field exactly (for example choosing a de 
Sitter background). That would give possibility to study the gravitational
stability of magnetic vacuum.

Second, there may be interesting cosmological applications of
this theory. In particular one may think of inflationary models without 
scalar fields where inflaton is composite vector.

Third, in realistic electroweak theory there is a stable
ground state of the type considered here. The effect of curvature
on the vacuum of this model will be considered elsewhere.

Finally, one may consider different Yang--Mills backgrounds,
(for example instantons) and discuss the gravitational effects there.
\bigskip
\centerline{\bf Acknowledgements}

S.D.O. would like to thank D. Amati and R. Iengo for the kind
hospitality at SISSA, where this work has been completed
and J. Ambj\o rn for useful remark.

\goodbreak
\centerline{\bf Appendix}
\smallskip
\noindent
It is interesting to compare the results of an SU(2) gauge theory 
to those of QED. For QED we find instead of eq.(17):
$$
L_{\rm eff}=-{1\over2}{e^2(0)\over e^2(t)}H^2
+{1\over 6}\left(a_3-{73 t\over 360(4\pi)^2}\right)R^2\ ,
$$
where
$$
e^2(t)={e^2(0)\over 1+{e^2(0) t\over 6\pi^2}}\ .
$$For zero curvature the maximum of the effective potential
$V_{\rm eff}=-L_{\rm eff}$ is unreliably large and
the minimum is for $H=0$. In general, however, curvature can influence
the position of both stationary points.

\goodbreak
\bigskip
\centerline{\bf References}
\smallskip
\item{1.} I.L. Buchbinder and S.D. Odintsov,
Izw. VUZov, Fiz. (Soviet Physics Journal) {\bf n4} (1983) 46;
{\it ibid.} {\bf n12} (1983) 108;
Yad. Fiz. (Sov. J. Nucl. Phys.) {\bf 40} (1984) 1338;
Lett. Nuovo Cim. {\bf 42} (1985) 379.
\item{2.} I.L. Buchbinder, S.D. Odintsov and I.L. Shapiro,
``Effective Action in Quantum Gravity'', IOP Publishing,
Bristol and Philadelphia (1992).
\item{3.} D. Gross and F. Wilczek,
Phys. Rev. Lett. {\bf 30} (1973) 1343;\hfil\break
D. Politzer,
Phys. Rev. Lett. {\bf 30} (1973) 1346.
\item{4.} G. 't Hooft, Nucl. Phys. {\bf B 62} (1973) 444.
\item{5.} G. Savvidy, Phys. Lett. {\bf B 71} (1977) 133;\hfil\break
S. Matinyan and G. Savvidy, Nucl. Phys. {\bf B 134} (1978) 539.
\item{6.} S. Coleman and E. Weinberg,
Phys. Rev. {\bf D 7} (1973)1888;\hfil\break
C. Ford, D.R.T. Jones, P.W. Stephenson and M. Einhorn,
Nucl. Phys. {\bf B 395} (1993) 17.
\item{7.} E. Elizalde and S.D. Odintsov,
Phys. Lett. {\bf B 303} (1993) 240;
{\it ibid.} {\bf B 321} (1994) 199;\hfil\break
E. Elizalde, A. Romeo and S.D. Odintsov,
Phys. Rev. {\bf D 51} (1995) 1680.
\item{8.} N.K. Nielsen and P.Olesen,
Nucl. Phys. {\bf B 144} (1978) 376;\hfil\break
J. Ambj\o rn, R.J. Hughes and N.K. Nielsen,
Ann. Phys. (NY) {\bf 150} (1983) 92;
H.B. Nielsen and P.Olesen,
Nucl. Phys. {\bf B 160} (1979) 380;\hfil\break
J. Ambj\o rn and R.J. Hughes,
Nucl. Phys. {\bf B 197} (1982) 113.
\item{9.} J. Ambj\o rn and P. Olesen, Nucl. Phys. {\bf B 330} (1990) 193.
\item{10.} E. Elizalde, 
Nucl. Phys. {\bf B 243} (1984) 398.
\item{11.} G. Shore, Ann. Phys. {\bf 128} (1980) 376.
\vfil
\eject
\bye